\documentclass[aps,prb,preprint,superscriptaddress,showpacs,floatfix]{revtex4}

\usepackage{graphicx}
\begin{document}

\title{Bias Dependent $1/f$ Conductivity Fluctuations in Low-Doped La$_{1-x}$Ca$_{x}$MnO$_3$ Manganite Single Crystals}

\author{M. Belogolovskii}
\affiliation{Donetsk Institute for Physics and Engineering, National Academy of Sciences of
Ukraine, 83114 Donetsk, Ukraine}

\author{G. Jung}
\author{V. Markovich}
\author{B. Dolgin}
\affiliation{Department of Physics, Ben Gurion University of the Negev, P.O. BOX 653, 84105 Beer
Sheva, Israel}

\author{X. D. Wu}
 \affiliation{Department of Physics, Ben Gurion University of the Negev, P.O.Box 653, 84105 Beer-Sheva, Israel}
 \affiliation{Department of Materials Engineering, Monash University, Clayton, Australia, 3800}

\author{Y. Yuzhelevski}
\affiliation{Department of Physics, Ben Gurion University of the Negev, P.O. BOX 653, 84105 Beer
Sheva, Israel}

\date{\today}

\begin{abstract}
Low frequency noise in current biased La$_{0.82}$Ca$_{0.18}$MnO$_{3}$ single crystals has been
investigated in a wide temperature range from 79~K to 290~K. Despite pronounced changes in magnetic
properties and dissipation mechanisms of the sample with changing temperature, the noise spectra
were found to be always of the $1/f$ type and their intensity (except the lowest temperature
studied) scaled as a square of the bias. At liquid nitrogen temperatures and under bias exceeding
some threshold value, the behavior of the noise deviates from the quasi-equilibrium modulation
noise and starts to depend in a non monotonic way on bias. It has been verified that the observed
noise obeys Dutta and Horn model of $1/f$ noise in solids. The appearance of nonequilibrium $1/f$ noise
and its dependence on bias have been associated with changes in the distribution of activation
energies in the underlying energy landscape. These changes have been correlated with bias induced
changes in the intrinsic tunneling mechanism dominating dissipation in
La$_{0.82}$Ca$_{0.18}$MnO$_{3}$ at low temperatures.
\end{abstract}

\maketitle

\section{Introduction}

Ultraslow relaxation in a variety of physical and nonphysical systems, manifesting itself in
$1/f^\alpha$ noise power spectral density (PSD) of the fluctuating observable quantity, with
$\alpha\approx 1$, continues to be one of the most exciting and puzzling effects attracting the
attention of many research groups.\cite{DH,weiss-rew} The most commonly investigated low frequency
fluctuations in solid state system are conductivity fluctuations. It is  generally recognized
that $1/f$ conductivity noise in solids arises from superposition of many elementary Lorentzian
contributions of individual  fluctuators with well defined and specifically distributed
characteristics relaxation rates.\cite{weiss-rew} Theoretical models of $1/f$ noise in solids
relate changes in the noise PSD to changes in the energy landscape in which charge carriers are
moving and dissipating energy. Therefore, studies of noise in solid state systems are important not
only from the point of view of improving the performance of solid state devices but also provide a
unique tool for understanding the nature and dynamics of transport processes in an investigated
system.\cite{weiss-tool}

The issue of noise in colossal magnetoresistance (CMR) manganites with general formulae
R$_{1-x}$A$_x$MnO$_3$,  where R is a rare-earth ion and A a divalent or tetravalent cation, was
addressed in many publications in the last decade. Almost all of them revealed prominent broad band
conductivity fluctuations with PSD following $1/f^\alpha$ law.  Low-doped La$_{1-x}$Ca$_x$MnO$_3$
(LCMO) system exhibits metal-to-insulator (M-I) transition in the vicinity of Curie temperature
$T_{\rm C}$. Above the M-I transition the conductivity of LCMO  is dominated by hopping mechanism
which strongly depends on oxygen stoichiometry and the tolerance factor. Sources of $1/f$ noise at
$T_{\rm C}$ were therefore initially associated with dynamics of oxygen vacancies.~\cite{raquet}

Early experiments concentrated on the noise peak associated with M-I transition, in the vicinity of
which a 3 to 4 order of magnitude increase of the noise level has been observed in Ca and Sr
doped manganites, see e.g.~\cite{raquet,raquetPRL,podz,alers,ahn}.  The noise peak around M-I transition has
been interpreted in terms of percolative nature of the transition between charge ordered insulating
and metallic ferromagnetic states,~\cite{podz} or in terms of magnetic fluctuations coupled to the
resistivity.~\cite{alers}  It became however soon evident that this unusually large $1/f$ noise is
not an intrinsic property of doped manganites. It was found that $1/f$ noise peak appears only in
notably strained films while it is completely absent in single crystals and in high quality, almost
strain-free, epitaxial films.~\cite{reut,palani,ourAPL} Small noise peak showing sometimes around
$T_{\rm C}$ in high quality samples could be easily suppressed by application of a small magnetic
field and was attributed to magnetic fluctuations in the vicinity of the phase transition.
Experimental observations of direct correlation between magnetic noise and magnetoresistance
allowed to conclude that the enhanced $1/f$ noise originates in charge carriers density
fluctuations.~\cite{rana} Current, temperature, and magnetic field dependence of noise in grain
boundary junctions, provided clear evidence that $1/f$ noise is caused by localized states with
fluctuating magnetic moments in heavily disordered grain boundary regions.~\cite{phillip}

Conductivity noise with $1/f$ spectrum is generally related to resistance fluctuations which are
measured by applying dc current and recorded as voltage fluctuations. When the resistance
fluctuations are just probed by current, and not influenced by its flow, then PSD of the noise
scales as the square of the bias current. Such modulation noise is  referred to as
quasi-equilibrium $1/f$ noise. There is a mounting experimental evidence that quasi-equilibrium
$1/f$ noise in doped manganites is accompanied by the noise that arises from, or is directly
modified by, the passage of current through the sample.~\cite{ourAPL,barone,nowak} This type of noise is referred to as nonequilibrium $1/f$ noise, and its dependence on bias is considerably different
from a quadratic one.

Our system of interest is La$_{0.82}$Ca$_{0.18}$MnO$_{3}$ (0.18 LCMO) single crystal, a highly
correlated electron system with a complex interplay of charge, orbital, and spin ordering, what
leads to remarkable changes in transport characteristics and magnetic state of the sample with
decreasing temperature. In this way 0.18 LCMO compound provides us with a unique opportunity of
studying noise properties in \emph{the same} solid-state sample, in markedly different transport
regimes and magnetic properties, using temperature and current bias as independent factors
modifying the state of the active fluctuators.

In this paper we demonstrate that equilibrium $1/f$ noise in La$_{0.82}$Ca$_{0.18}$MnO$_3$ single
crystals preserves its character under changing temperature, despite significant changes in magnetic and transport properties of the system. Bias dependent nonequilibrium $1/f$ fluctuations appear
only at low temperatures, well below the Curie temperature. Properties of the nonequilibrium $1/f$
noise differ significantly from those of quasi-equilibrium one. A combined analysis of noise and
transport characteristics permits us to associate marked changes in the noise behavior to changes
in the low temperature intrinsic tunneling mechanism.

\section{Experimental background}

La$_{0.82}$Ca$_{0.18}$MnO$_3$ crystals were grown by a floating zone method using radiative
heating.\cite{crystal} The crystallographic orientations of the crystal and wafers were determined
by Laue method with accuracy of $\pm 2^{\circ}$. X-ray data of the crystal were compatible with the
perovskite structure orthorhombic unit cell, $a = 5.5062$ \AA,  $b = 7.7774$ \AA, $c = 5.514$ \AA.
The as-grown crystal, in form of a cylinder, about 4 cm long and 4 mm in diameter, was cut into
individual small rectangular 6 $\times$ 3 $\times $ 2 mm$^3$ bars, with the longest dimension along
the $<110>$ crystallographic direction. Current and voltage leads were indium soldered to gold/chromium
contacts deposited by thermal evaporation in vacuum.

For noise measurements the sample was thermally anchored to the sample holder of a variable
temperature liquid nitrogen cryostat. Conductivity noise was measured in a conventional 4-point
contact arrangement  by biasing the sample with dc current supplied by high output impedance
current source and measuring the resulting voltage fluctuations. Four in-line contacts were placed along the longest dimension of the bar. The separation between voltage
contacts was 0.3 mm. The voltage signal was amplified by a home made room temperature low noise
preamplifier, located at the top of the cryostat, and further processed by a computer assisted
digital signal analyzer. To eliminate environmental interferences and noise contributed by the
measuring chain, the PSD measured at zero current was subtracted from the data obtained at a given
current flow for each measurement.

\begin{figure}
\includegraphics*[width=12truecm]{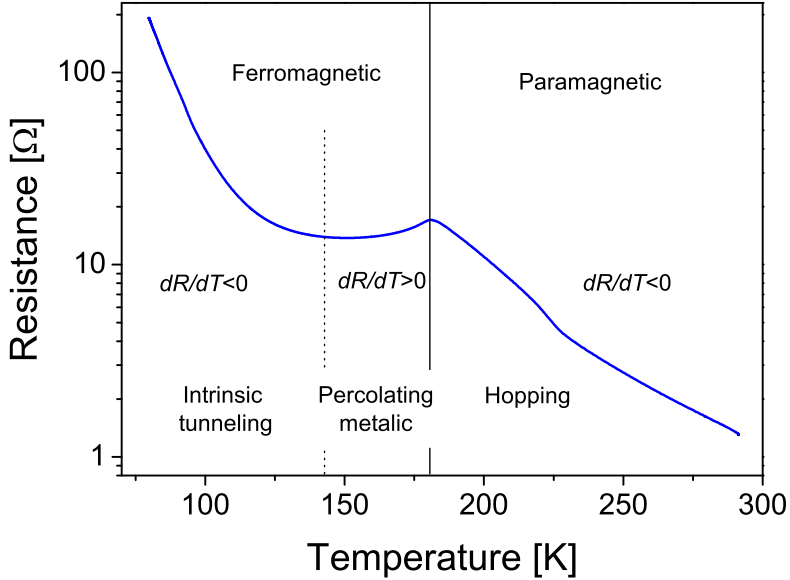}
\caption{Color online. Temperature dependence of the resistance of La$_{0.82}$Ca$_{0.18}$MnO$_3$ crystal at zero
applied magnetic field. The room temperature resistivity of the crystal is 2.4 $\Omega cm$.} \label{RT}
\end{figure}

Remarkable changes in transport properties and magnetic state of 0.18 LCMO system upon changing
temperature are illustrated in Fig. \ref{RT}. At high temperatures the sample is in paramagnetic
insulating state and the resistivity, dominated by hopping mechanism, increases with decreasing
temperature. The resistivity reaches a pronounced maximum related to the metal-insulator (M-I)
transition at $T=T_{\rm {MI}}=180$ K. The maximum appears at temperature very close to the Curie
temperature $T_{\rm C}$ of the paramagnetic-to-ferromagnetic transition. $T_{\rm C}$ determined by
independent magnetization measurements is $180 \pm 1$ K. Intrinsic phase separation
(PS) associated with metal-insulator (M-I) transition at $T\sim T_{\rm C}$ leads to percolation
conductivity in the ferromagnetic (FM) state at $T< T_{\rm C}$. Therefore, at temperatures below
the magnetic ordering temperature the resistivity decreases in a metallic-like way with $dR/dT >
0$, although its absolute value is much higher than that of common metals. With further temperature
decrease, the resistivity reaches a shallow minimum around $T\sim 120$ K, followed by a strong
upturn at temperatures below some 100 K.

Detailed magnetic measurements of our sample did not reveal any peculiarities at temperatures below
$T_{\rm C}$, suggesting that in the investigated temperature range there is only one magnetic
transition at $T=T_{\rm C}$. Low temperature resistivity is dominated by tunneling across intrinsic
barriers associated with extended structural defects, such as twins and grain boundaries, and/or
with inclusions of insulating FM phase interrupting metallic percolating paths.\cite{PRB018}

\begin{figure}
\includegraphics*[width=12truecm]{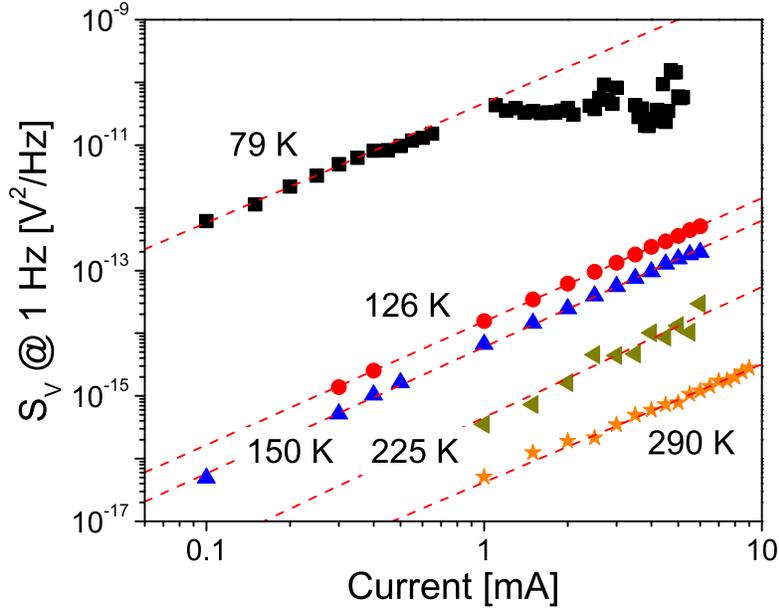}
\caption{Color online. Current dependence of the noise intensity at $f=1$ Hz as recorded at various temperatures.}
\label{PSD(I)}
\end{figure}

Despite changes in the dissipation mechanism and magnetic state of the sample, the experimentally
observed voltage noise was found to have $1/f$ spectrum in the entire investigated temperature and
bias range. To get a first insight into the nature of the observed noise, we have measured the
dependence of the noise intensity on bias current. We found that for all temperatures, with the
exception of the 79 K data, for currents exceeding $I\sim $1 mA, the noise intensity scales as the
square of the bias current, as illustrated in Fig. \ref{PSD(I)}. Proportionality of  $1/f$ voltage
noise to the current squared indicates that the noise is due to bias independent resistivity
fluctuations which are unaffected by the current flow.

At low temperatures, where electrical transport in our sample is dominated by tunneling
mechanism,\cite{PRB018} at currents exceeding some threshold current $I\sim$ 1 mA, the
proportionality of the voltage noise to $I^2$ breaks down and the resistivity fluctuations  start
to be influenced by the current. This is a manifestation of a nonequilibrium $1/f$ noise similar to
that frequently observed in  nonlinear, nonohmic systems.~\cite{zygulski} However, in a difference
to other nonlinear systems, the crossover to nonequilibrium noise in our sample is not associated
with the onset of strong nonlinearity in the current-voltage $(I-V)$ characteristics. \cite{nowak} Moreover, the
intensity of nonequilibrium noise initially decreases with increasing bias, to change in a
nonmonotonic way with further increase of the bias.

The puzzling properties of the low temperature noise in La$_{0.82}$Ca$_{0.18}$MnO$_3$ single
crystalline sample should be related to the underlying changes in its transport properties. An
efficient way to approach the problem is to detect possible changes in the energy environment of
the fluctuators responsible for the noise. A tool for such analysis is provided by relevant
theoretical models of nonexponential relaxation kinetics resulting in $1/f$
noise.~\cite{DH,weiss-rew}

\section{Theoretical background}

\subsection{1/f noise in solids}

$1/f$ noise spectra in condensed-matter systems are believed to come from assembly of elementary
fluctuators with well defined characteristic relaxation rates $\lambda$. The spectrum of each
elementary fluctuator is Lorentzian, and the resulting normalized PSD of the fluctuating quantity
$X$,
\begin{equation} S(f)=S_X(f)/X^2\propto \int{\frac{\lambda }{\lambda^2+f^2}P(\lambda
)}d\lambda. \label{Lor}
\end{equation}
will have $1/f^\alpha$ form, with $\alpha = 1$ for the distribution function $P(\lambda )\propto
1/\lambda$. If the elementary fluctuators are thermally activated, the characteristic rate $\lambda
(\Delta)=\lambda_0\exp(-\Delta/k_{\rm B}T)$, and the required distribution of $\lambda$ is provided
by flat distribution of activation energies, $P(\Delta)=const$.

Dutta and Horn, (DH) have shown that $1/f^\alpha$ spectra with $\alpha\approx 1$ arise not only for
$P(\Delta)=const.$ but also for the distribution function $P(\Delta )$ that does not vary much in
the range of $\Delta $ from $|k_{\rm{B}}T\ln(\lambda_0/f_1)|$ till $|k_{\rm B}T\ln(\lambda_0/
f_2)|$, where $f_1$ and $f_2$ are the lowest and the highest frequency of the
measurements~\cite{DH}. For a detailed discussion of DH model assumptions see reference
~\onlinecite{weiss-rew}.

When $P(\Delta )$ is a slowly varying function of the activation energy $\Delta $ then
\begin{equation}
S(f,T)f\propto k_{\rm B}TP(\tilde{\Delta}=k_{\rm B}T\ln(\lambda _0/f)). \label{PSD-D}
\end{equation}

In the majority of solid state systems $\lambda _0$ is a constant of the order of phonon
frequencies $\lambda _0\sim 10^{12}-10^{13}$ Hz.~\cite{DH} But if a disordered system undergoes a
phase transition at transition temperature $T_{\rm C}$, then for the noise spectra in the vicinity
of $T_{\rm C}$ we obtain the same Eq.~\ref{PSD-D} but with $k_{\rm B}T$ replaced by $1/\xi
^{\theta}$, where $\xi$ is the correlation length and $\theta $ is a dynamic critical
exponent.~\cite{imry} As a result, while analyzing experimental data using Eq.~\ref{PSD-D} one may
obtain unreasonably large values of $\lambda _0$.

In the experiments we have measured the noise in the frequency range from $f_0=1$ Hz to $f=10^2$ Hz. Therefore, $\ln
(f/f_0)\leq 5$ may be considered as a small addition to $\ln(\lambda _0/f_0)\approx 25-30$ and we
can further expand the $P(\tilde{\Delta}=k_{\rm B}T\ln(\lambda_0/f))$ function from the right-hand side of
Eq.~(\ref{PSD-D}) as
\begin{equation}
P(k_{\rm B}T\ln (\lambda _0/f))\approx P(k_{\rm B}T\ln(\lambda _0/f_0))- \frac{dP}{d\Delta
}\bigg\vert_{\breve{\Delta}=k_{\rm B}T\ln(\lambda _0/f_0)}k_{\rm B}T\ln(f/f_0). \label{lnf}
\end{equation}
Within the DH approximation, at low frequencies,  $S(f,T)f$-vs.-$\ln (f/f_0)$ dependence should be
linear at a fixed temperature. It follows from Eqs. \ref{PSD-D} and \ref{lnf} that the slope of
$S(f,T)f$-vs.-$\ln (f/f_0)$ plot is proportional to $d P(\Delta )/d \Delta $. This slope is directly
related to PSD exponent $\alpha$ through a general relation
\begin{equation}
\alpha (f,T)=1-\frac{1}{S(f,T)f}\frac{\partial (S(f,T)f)}{\partial\ln f}.
\label{DHT-alpha}
\end{equation}

DH model leads to a specific relation between temperature derivative of the spectral density
$\partial \ln (S(f,T)f)/\partial \ln T$ and the derivative $\partial \ln (S(f,T)f)/\partial \ln f$
\begin{equation}
\frac{\partial \ln (S(f,T)f)}{\partial \ln T}=1- \ln(\lambda
_0/f)\frac{\partial \ln (S(f,T)f)}{\partial \ln f}. \label{DH}
\end{equation}
Equation (\ref{DH}) contains crucial reciprocity between the frequency and temperature dependence
of the noise magnitude and is frequently used as a self-consistency test for the validity of DH
approach for a given physical system. The relevant test consists in verifying whether the
dependence of the noise intensity on frequency at a constant temperature, and on temperature at a
constant frequency, are consistent with each other. If this is the case, one may use Eq.
(\ref{PSD-D}) to attribute a single distribution of activation energies to the ensemble of active
fluctuators. Flat distribution of temperature independent activation energies in DH model gives
rise to a pure $1/f$ spectrum and a linear temperature dependence of the noise level at low
frequencies. Departures of $\alpha$ from unity indicate non-zero derivatives of energy distribution
$dP(\Delta)/d\Delta$, and excess of high ($\alpha <1$), or low energy ($\alpha
>1$) fluctuators in the ensemble.

\subsection {Inelastic transport across an inhomogeneous medium.}

At lowest temperatures studied in the experiment, charges in the investigated system are
transmitted by tunneling across classically forbidden regions. Electron wave functions in tunneling
barriers decay exponentially over the characteristic length $l$. When the decay length $l$ exceeds
the barrier thickness $d$, then direct tunneling, both elastic and inelastic, dominates the
conduction mechanism. Elastic contribution to the differential conductance
$G(V)=dI(V)/dV$-versus-voltage characteristic is a parabola~\cite{Wolf}, with
\begin{equation}
G_{{\rm{el}}}(V) = G_0  + {\rm{const}} \cdot (V-V_0)^2,
\end{equation}
where $G_0$ is proportional to $\exp(-2d/l)$.

The inelastic contribution to the differential conductance is given by $G_{\rm{in}}(V) \propto
\int\limits_0^{eV} {a^2 F(\omega )d\omega }$ with the Eliashberg electron-boson interaction
function $a^2 F(\omega )$.~\cite{Wolf} For most metals, in the case of phonons as mediating bosons,
the shape of the average Eliashberg spectral function resembles that of the phonon density of
states $F(\omega )$.~\cite{Maximov} Let us look, for example, at $F(\omega )$ obtained in neutron
scattering experiments for a polycrystalline sample of
La$_{0.625}$Ca$_{0.375}$MnO$_3$.~\cite{Adams} One can see that, within the range of 10 to 70 meV,
the average deviation of the phonon density of states from its mean value does not exceed 20 \%.
Therefore, in this energy interval one may use the approximation $F(\omega )=\rm{const}$, which
results in a linear behavior of $G_{\rm{in}}$ vs. $V$; $G_{\rm{in}}(V)\propto
|V|$.~\cite{kirtley} Such situation has been frequently observed in disordered perovskite oxides,
high $T_{\rm C}$ cuprates and various CMR manganites. ~\cite{bel}

With increasing thickness of the tunneling  barrier, hopping along chains of localized states is
favored. While hopping along localized states path, a carrier does not cross quantum-mechanically
the entire distance between the electrodes, but rather jumps from the junction electrode to the
first state, lose the phase memory,  moves to the second nano-island and, eventually, after
completing all the hopping path, jumps to the opposite electrode. Such case was considered by
Glazman and Matveev~\cite{GM,Beasley} in their model (GM) of indirect tunneling in disordered
materials.  The GM model applies very well to low temperature transport in perovskite manganites.~
\cite{PRB018,gross,bertina}

In the GM model, temperature and voltage dependencies of the tunnel conductance are expressed as
multistep tunneling via $N$ localized states:
\begin{equation}
G(V)=G_0+\sum_{N=1}^{\infty}G_N(V,T), \label{14}
\end{equation}
where conductance $G_0$ represents bias and temperature independent elastic tunneling term, while
$G_N$ describe tunneling through $N\geq1$ localized states.
\begin{eqnarray}
G_N(V)=a_NV^{(N-\frac{2}{N+1})} \hskip 1truecm\mbox{for}\hskip 0.3truecm eV\gg k_BT,\\
G_N(T)=b_NT^{(N-\frac{2}{N+1})} \hskip 1truecm\mbox{for}\hskip 0.3truecm eV\ll k_BT, \label{Gn}
\end{eqnarray}
where coefficients $a_N$ and $b_N$ depend exponentially on barrier thickness.

Resuming, the differential conductance of a thin, inhomogeneous, insulating barrier is a power
function of the voltage, see also Ref.~\onlinecite{CEJP},
\begin{equation}
G_{\rm{in}}(V) = G_0+{\rm{const}}\cdot V^n \label{power},
\end{equation}
where index $n$ characterizes the tunneling regime: $n=2$ corresponds to elastic tunneling with the
energy relaxation in the conducting regions of the system, whereas other $n$ are signatures of
inelastic tunneling in which an electron losses its energy inside the insulating region. By finding
the value of index $n$ from experimental data one can infer information about the physics of
electron transport processes across the dielectric layer.

\section{Results and discussion}

\subsection{Applicability of DH model}
\begin{figure}
\includegraphics*[width=12truecm]{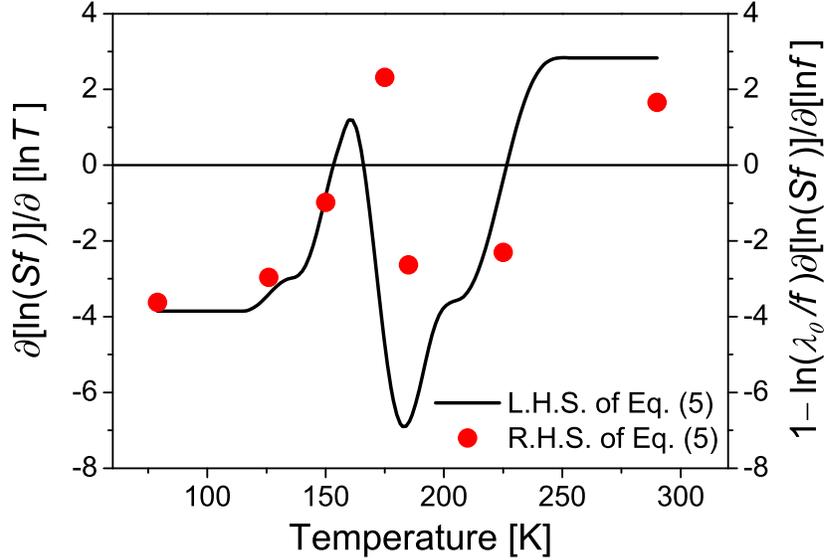}
\caption{Color online. Verification of the DH temperature-frequency reciprocity.
Comparison of the left- and right-hand sides of Eq.~(\ref{DH}) at
$f=2$ Hz and $I=1.5$ mA.} \label{verDHD}
\end{figure}

It should be noticed that deviations or inconsistencies with DH model in solid state systems are
rare.\cite{weiss-rew}  In particular, it has been already demonstrated that noise in various
half-metallic oxides obeys the DH model.\cite{raquet} Nevertheless, as a first step we perform
tests of the applicability of DH model to our data by checking the reciprocity condition expressed
in Eq.~(\ref{DH}) which relates the temperature and frequency derivatives of the dimensionless PSD
at fixed $T$ and $f$ values. Figure ~\ref{verDHD} compares left and right sides of Eq.~(\ref{DH})
for frequency equal to 2 Hz, current bias of 1.5 mA and all temperatures studied. The overall
agreement is good, including also the 79 K data, and even the change of the sign around 175 K is
well reproduced in the both dependencies.

Canonical DH approach assumes that $P(\Delta)$ is temperature independent. We find a small violation of this assumption for temperatures around Curie temperature $T_C$. The parameter $\lambda _0$ estimated from the best agreement, in the least-squares sense, between two dependencies plotted in Fig.~\ref{verDHD} was found to be $\lambda _0=4.3 \times 10^{14}$ Hz. Thus obtained $\lambda _0$ is physically reasonable, although it is about two orders of magnitude higher than a typical phonon frequency. We attribute this discrepancy to the existence of the phase transition in the investigated system at $T=180$ K.~\cite{weiss-rew,imry} Indeed, separate estimations of $\lambda _0$ for each individual temperature render values close to the expected phonon frequency at high and low temperatures, while values of $\lambda _0$ for temperatures close to $T_C$ are unphysical and diverge.

We conclude that the frequency-temperature dependence reciprocity condition is fulfilled and that
the observed noise does obey the DH model in entire investigated temperature range. We are allowed
therefore, to assume that the observed noise is due to thermally activated kinetics of active
fluctuators with the distribution of activation energies $P(\Delta)$ given by Eq. (\ref{PSD-D}),
and the derivative of activation energy distribution $d P(\Delta )/d \Delta $ proportional to the
slope of $S(f,T)f/V^2$-vs.-$\ln (f/f_0)$ characteristics.

\subsection{Quasi-equilibrium noise.}

\begin{figure}
\includegraphics*[width=12truecm]{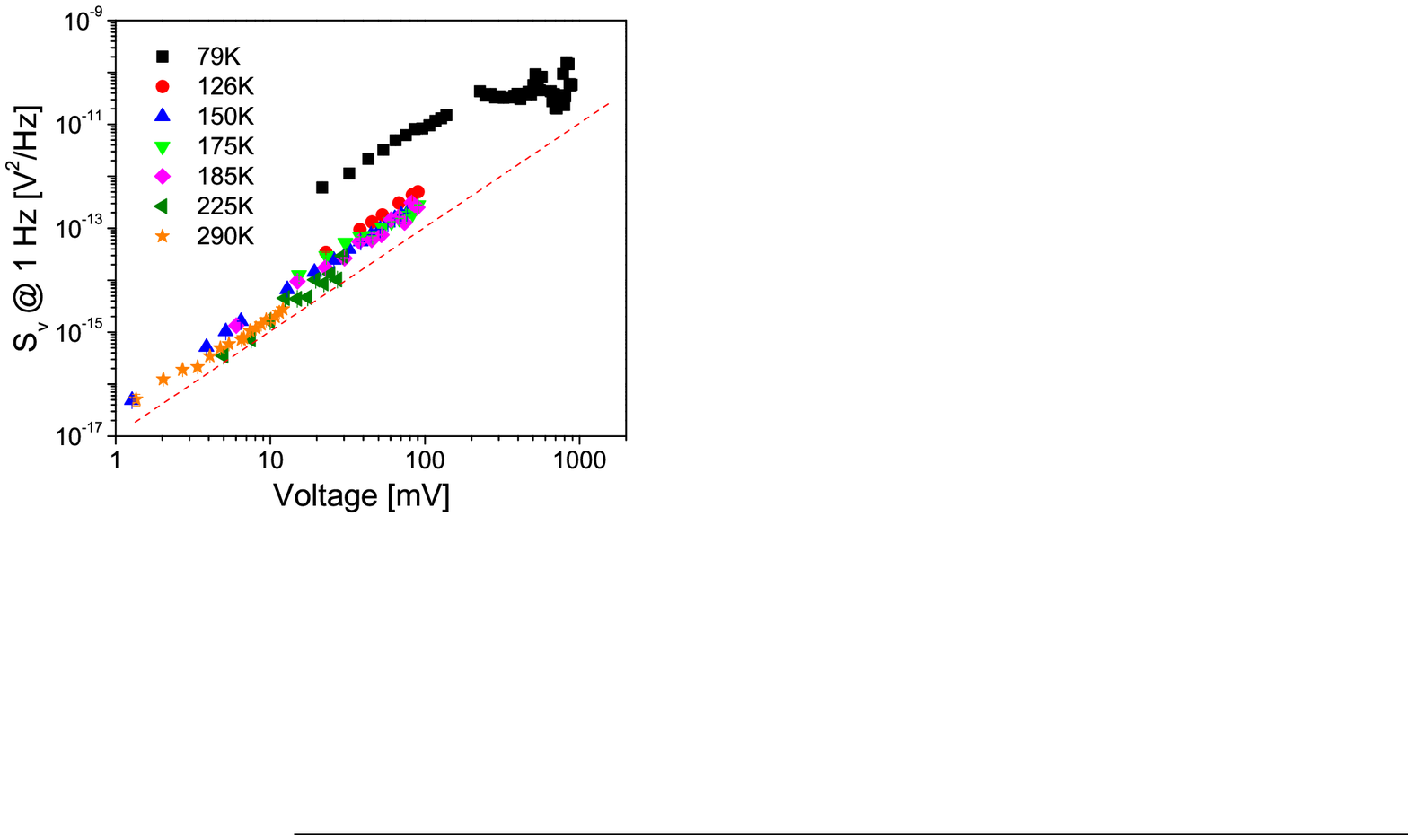}
\caption{Color online. Voltage dependence of the noise intensity at $f=1$ Hz as recorded at various temperatures.} \label{SvvsV}
\end{figure}

Figure \ref{SvvsV} shows the data from Fig.~\ref{PSD(I)} replotted as a function of bias voltage.
Within the experimental accuracy data recorded at all temperatures, with the exception of $T=79$ K,
collapse into single line $S_V\propto V^2$. This is equivalent to the statement that
$S_V/V^2=const$, meaning that the reduced spectral density of resistance fluctuation
$S_r=S_R/R^2=S_V/V^2$ is temperature independent. Temperature independent $S_r$ implies that the
dependence of resistance fluctuations $\delta R$ on temperature follows the temperature dependence
of the resistance of the sample $R$. Experimentally revealed condition $\delta R\propto R$ assures
as well that the coupling of the fluctuators to resistance fluctuations is temperature independent.
Observe that temperature independence of the coupling of system fluctuations to the measured
physical parameter is one of the most problematic condition assumed in the Dutta-Horn model of
$1/f$ noise and its violation frequently leads to some deviations of the experimental data from the
simplest version of the model.\cite{weiss-rew} On the other hand the temperature independence of
the above coupling is quite surprising result if one takes into account pronounced changes of the
magnetic state and dissipation mechanism in the sample with changing temperature.

\begin{figure}
\includegraphics*[width=12truecm]{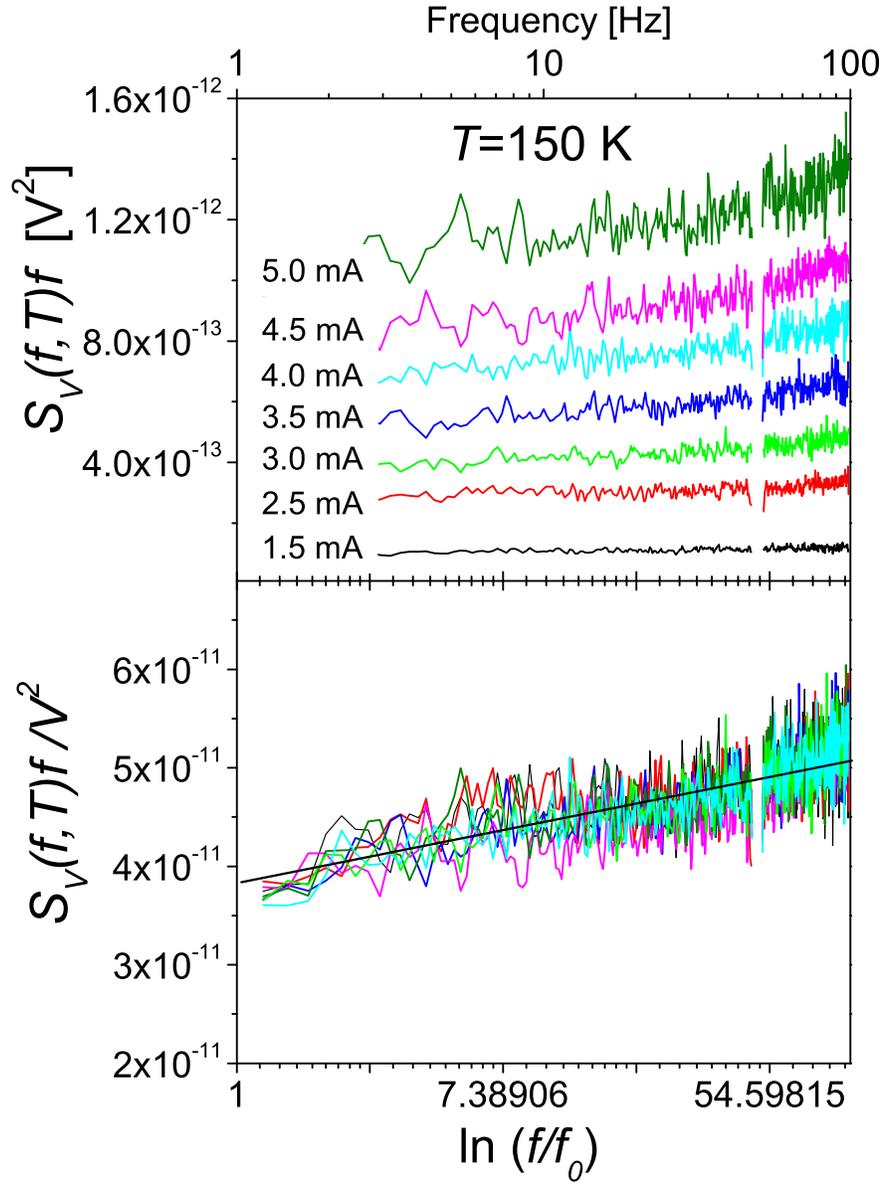}
\caption{Color online. Upper panel: Example of a linear dependence of the product $S_V(f,T)f$ as a function of $\ln f/f_0$ recorded for various currents at $T=150$ K. Lower panel: Data from the upper panel after normalization of the
noise intensity by $V^2$. Observe that all data from the upper panel collapse to a single line
after the noise normalization. Solid straight line is a linear-fit approximation of the
experimental data for all current biases.} \label{verapprox150KK}
\end{figure}

\begin{figure}
\includegraphics*[width=12truecm]{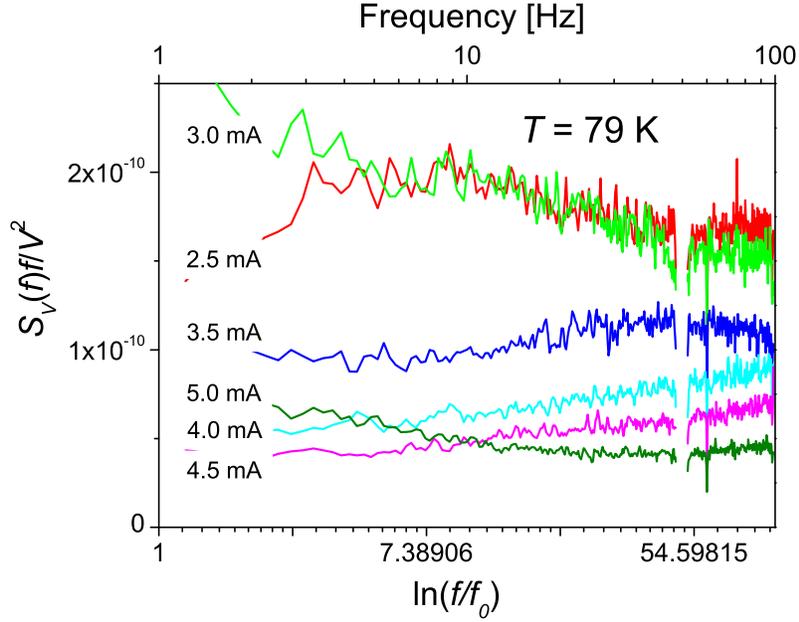}
\caption{Color online. Effect of the current bias on the dimensionless noise
intensity recorded at 79 K.} \label{verapprox79K}
\end{figure}

Upper panel of Figure~\ref{verapprox150KK} shows $S(f,T)f$  data recorded at 150 K for different
bias currents and plotted as a function of $\ln (f/f_0)$, while the lower panel shows the same
data normalized by dc voltage squared, $S(f,T)f/V^2$.
First of all, one observes that $S_V(f,T)f$ vs. $\ln f/f_0$ at a fixed  temperature is indeed a linear function as predicted by Eq.~(\ref{lnf}) (similar results were obtained for other currents as well). Moreover, all normalized spectra
recorded at different currents collapse to a single line, consistently with the $I^2$ dependence of
the equilibrium noise. The conductivity noise at 150 K  is thus only probed and not modified by
current flow. By fitting the normalized data to a linear function (a solid straight line in
Fig.~\ref{verapprox150KK}) one obtains the derivative $\partial (S(f,T)f)/\partial \ln f$.
Therefore, we may conclude immediately that at $T=150$ K bias current $I$ has no influence
whatsoever on the distribution of activation energies $P(E)$ since the slope of each spectrum is
the same within the experimental accuracy, meaning that $d P(\Delta )/d \Delta $ does not depend on
bias. The same independence of the slope factor, and consequently $P(\Delta)$ function, on current
has been confirmed for all temperatures with exception of 79 K data. Here, even a brief look at the
plot $S(f,T)f/V^2$-vs-$\ln f$ from Fig. \ref{verapprox79K} proves that at 79 K, in a marked
difference to higher temperatures, the distribution function and its derivative are strongly
influenced by the current flow. This is reflected in a strong dependence of the spectral exponent
$\alpha$ on current flow at 79 K, see Eq.~\ref{DHT-alpha}, in contrast to a very weak current
influence on $\alpha$ at higher temperatures.

\subsection {Nonequilibrium noise.}

In order to measure resistance fluctuations in the equilibrium state, a small current is applied to
a sample with the only aim to convert the resistivity fluctuations into measurable voltage noise.
In this case the dimensionless PSD spectrum $S_{\rm V}(f,T)f/V^2$ is a constant which does not
depend on  voltage, or current, bias.

It can be clearly seen in Figs. \ref{PSD(I)} and \ref{SvvsV}, that $1/f$ noise at 79 K looses its
equilibrium character at bias exceeding 1 mA, or equivalently, 200 mV. Above the threshold bias the noise is no
more proportional to bias squared. The nonequilibrium noise intensity changes in a nonmonotonic
way with increasing bias and decreases with increasing bias at certain bias ranges.

There are several reasons why it may happen. We shall discuss here four possibilities which can
have direct relation  to our study. The first two are artifacts associated with shunting parts of the sample under the voltage contacts by the contact pads or  simple overheating of a sample due to dissipation at
high current biases. The third possibility is a direct influence of the external bias on
elementary fluctuators. Let us consider a single two-state system with equal activation energies of
both states, symmetric two well energy structure, at zero bias. If the energy structure of the
fluctuator is stressed by the bias, the energy difference between metastable states 1 and 2  will increase with increasing bias. The total switching rate is a sum of the two
switching rates $\lambda_{1,2}=\lambda_0\exp(-\Delta _{1,2}/k_{\rm B}T)$ for two metastable
states.~\cite{Machlup} This mechanism may lead to increasing asymmetry in the energy structure of
the involved fluctuators and, consequently, in  a decrease of  noise intensity with increasing
bias. A fourth possibility  of remarkable bias enforced modifications of the noise PSD, which would
be most important for inhomogeneous samples, are changes in the mechanisms of electron transmission
through the investigated sample.

Let us first convince the reader that the observed nonequillibrium noise is not an artifact resulting simply from contacts influence or overheating of the sample by flowing current, as indicated above. At low temperatures the resistivity of the LCMO crystal increases by more than an order of magnitude. Therefore, voltage contacts shunting effects may become significant, in particular for contacts separated by distances comparable, or smaller than the sample thickness, as in our case. However, one expects shunting to manifest itself as a weaker than quadratic dependence of the noise power on bias, but not as a decrease of the noise with increasing bias, as in our experiments.  The decisive  test for contacts effects consisted  in comparing the results of standard 4-point measurements with data obtained in an inverted contact arrangement, in which current was fed into the sample through inside pair of voltage contact and voltage noise measured at outside current contacts. The bias at which the onset of nonequillibrium noise appears in the inverted contact arrangement was found to be the same as the onset bias measured using standard contact arrangement. Moreover, in early experiments we have used much bigger, 2.6 mm separation between the voltage contacts.  The measured value of low temperature resistivity of our sample was not dependent on the distance between the voltage contacts. Together, the above observations enabled us to reject the possibility that low temperature noise behavior is associated with contact artifacts.

The increase of the resistance by more than an order of magnitude leads to increase of the the power dissipated in the sample with respect to the power dissipated at higher temperatures at the same current flow. This is clearly seen in Fig.~\ref{SvvsV} where the nonequillibrium noise at 79 K appears at bias voltages, and power levels, significantly exceeding those reached at higher temperatures. Temperature independence of the normalized spectral density of resistivity fluctuations $S_r$
causes the  measured noise to follow the temperature evolution of the sample resistance squared. At
temperatures below $\approx 100$ K the resistance decreases with increasing temperature, see Fig.
\ref{RT}. Therefore, one may suspect that  decrease of the noise at 79 K for currents above 1 mA
and deviations from the equilibrium quadratic scaling may be due to simple overheating by current
flow. Indeed, it is possible to explain qualitatively the noise behavior if one assumes that
current flow above 1 mA increases the temperature of the sample, thus causing the resistivity
decrease, what should lead to the  decrease of the noise. This scenario requires however huge
overheating of the sample by almost 100 K. This is unlikely in our experimental arrangement.
Moreover, the changes in sample resistance measured at 79 K as a function of current are negligible
with respect to those required to explain the noise behavior by overheating. Therefore, a simple
overheating scenario outlined above cannot explain our data.

\begin{figure}
\includegraphics*[width=12truecm]{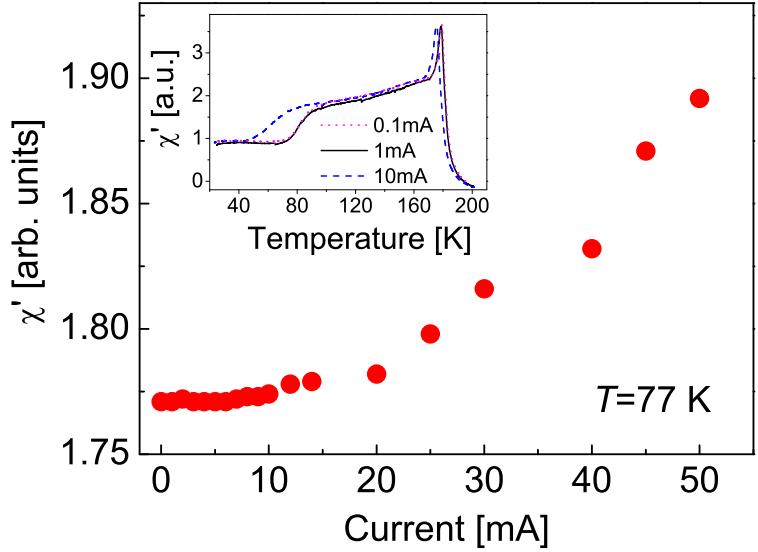}
\caption{Color online. Real part of magnetic susceptibility as a function of the
current bias. Inset shows temperature dependence of susceptibility
with $I$=0, 0.1, and 10 mA current flow in the sample.}
\label{chi}
\end{figure}

Nevertheless, to completely reject the plausibility of the overheating scenario we have performed
independent measurements of the sample magnetic susceptibility under current flow. Figure~\ref{chi}
shows the real part of magnetic susceptibility  as a function of dc current, while the insert
demonstrates the temperature dependence of susceptibility  measured at different currents. It is
evident that at $T=$77 K there is noticeable increase of the sample temperature for currents above
$I = 8$ mA. This current level is almost an order of magnitude higher than the threshold current at
which nonequilibrium noise appears. Therefore, independent measurements of temperature and current
dependence of magnetic susceptibility enable us to fully reject the overheating scenario.

To find the real origin of the noise evolution with increasing bias at 79 K, we have performed the measurements of the voltage dependence of the differential conductance $G(V)$ and compared it with Eq.~\ref{power}. We have determined the conductance exponent $n$ by fitting the $G(V)$ characteristics to Eq.~\ref{power} in several voltage ranges and verified the results by differentiating the experimental curves $G(V)$ and calculating the value of $n$ as
\begin{equation}
n=1+\frac{d\ln[\frac{dG(V)}{dV}]}{d\ln [V]} \label{calcn}
\end{equation}

\begin{figure}
\includegraphics*[width=12truecm]{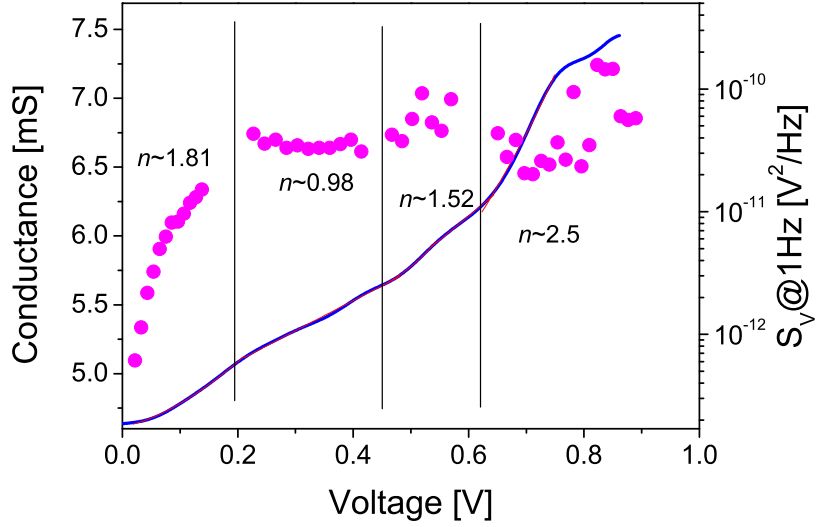}
\caption{Color online. Effect of the voltage bias on the differential
conductance (a solid line) and the normalized noise spectrum at
frequency 1 Hz (circles) recorded at 79 K. Vertical lines show
voltage intervals where the exponent $n$ in Eq.~\ref{calcn} are
nearly constant, corresponding values of $n$ are indicated in the
figure.} \label{SvdVdIn-vs-V}
\end{figure}

Figure \ref{SvdVdIn-vs-V} shows details of the dependence of $1/f$ noise intensity and sample conductance on bias voltage at 79 K, together with the evaluated values of exponents $n$. We find that for voltages below 195 mV $n=1.81 \pm 0.02$. Most probably, the dominating tunneling mechanism in this voltage range is direct elastic tunneling across the insulating interlayer between the conducting banks, see the previous section. For the bias range between the two first maxima of the noise, i.e. between 195 and 450 mV, the dependence $G(V)$ is almost linear with $n=0.98 \pm 0.01$. It has been argued~\cite{kirtley} that a linear conduction background with $n=1$ appears due to strong direct inelastic tunneling involving a broad continuum density of states of bosonic excitations inside the tunneling barrier. This situation has been frequently experimentally observed in disordered perovskite materials like high-$T_{\rm c}$ superconductors and various CMR manganites.~\cite{bel}

For voltages between 450 mV to 620 mV, i.e., above the second peak of the noise, we find that
$G(V)$ is composed  of two shifted nonlinear pieces, each with $n=1.36 \pm 0.02$, giving the
overall exponent in this range $n=1.52\pm 0.02$. At voltages between 620 and 870 mV, i.e., above
the second peak, we find $n=2.5 \pm 0.03$. The two last voltage ranges correspond to indirect
inelastic tunneling via more than two conducting inclusions inside the insulating
interlayer.~\cite{GM}

Changes of the exponent $n$ reflect changes in the conduction mechanism. As we have discussed in
details previously,  the electron transport in La$_{0.82}$Ca$_{0.18}$MnO$_3$ at low temperatures is
dominated by intrinsic tunneling mechanism.~\cite{PRB018} It is therefore tempting to analyze the
behavior of the exponent $n$ in terms of changing tunneling mechanisms. This approach encounters,
however, formal difficulty associated with the fact that our sample does not constitute a well
defined single tunnel junction. The tunneling conductance in bulk La$_{0.82}$Ca$_{0.18}$MnO$_3$ is
associated with intrinsic tunnel barriers appearing spontaneously along the ferromagnetic metallic
percolation paths.\cite{PRB018} The exact nature and individual properties of the barriers cannot
be characterized as one has an experimental access only to some averaged ensemble-like tunnel
characteristic of the sample. Therefore, we can only approximate the characteristics of intrinsic
junctions with the known characteristics of well defined discrete junctions.

The scenario in which the noise intensity varies with changing bias due to changes in conduction
mechanism can be outlined as follows. At low bias voltages a direct tunneling mechanism is
dominating, $n$ is close to 2 for dominantly elastic charge transmissions and close to 1 in the inelastic
tunneling regime. Crossover to nonequilibrium noise is associated with the change of $n$ from 2 to
1, i.e., from elastic to inelastic transmission, which occurs when $\alpha=1$. At higher voltages
the probability of indirect inelastic tunneling is much higher since the number of involved
localized states in the barrier grows with increasing bias.~\cite{Beasley} The conductivity of an
inelastic channel increases exponentially with increasing number $N$ of the localized states in the barrier.
Therefore, the current is shunted by channels with higher $N$ whenever available. The
experimentally determined exponent $n=1.5$ corresponds to admixture of indirect tunneling
through channels with two ($n=1.3$) and three ($n=2.5$) localized states in the barrier. A similar mechanism has been invoked previously to explain unusual decrease of the noise with increasing bias in magnetic tunnel junctions. \cite{nowak}

Figure~{\ref{alpha-V_79} shows how the spectral exponent $\alpha $ at 79 K behaves with increasing
voltage bias. The exponent reflects directly changes in the derivative $\partial (S(f,T)f)/\partial
\ln f$ obtained from the slope of $S(f)f$-vs-ln $f/f_0$ characteristics and  related to $\alpha$ by
Eq.~(\ref{DHT-alpha}). Figures \ref{SvdVdIn-vs-V} and \ref{alpha-V_79} evidence apparent
correlations between the behavior of the noise, evolution of the exponent $\alpha$, and
corresponding $n$ values. It seems that significant changes in the difference $\delta \alpha
=\alpha-1$, or equivalently the changes of the sign of the derivative $\partial (S(f,T)f)/\partial
\ln f$, appear in close vicinity of voltages for which $\delta \alpha =0$, i.e. for voltages at which the
distribution function $P(\Delta )$ is a constant independent on the magnitude of an activation
energy.

Changes of the conduction mechanism, reflected by changes in $n$, are accompanied
by clear noise peaks. The peaks can be seen as manifestation of excess
partition noise  at specific voltages at which alternative inelastic channels with different $N$ are
available for electron transfer. We conclude that Figs. \ref{SvdVdIn-vs-V} and \ref{alpha-V_79}
evidence a clear correlation between bias induced changes in the noise and modifications of $I-V$
characteristics reflecting changes in the intrinsic tunneling mechanism.
\begin{figure}
\includegraphics*[width=12truecm]{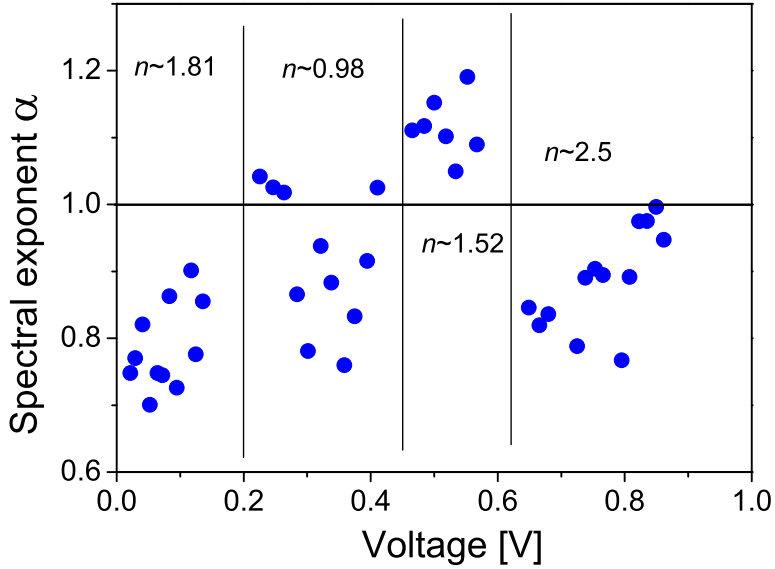}
\caption{Color online. Effect of the voltage bias on the parameter $\alpha $ at
frequency 1 Hz (circles) recorded at 79 K. Vertical lines show
voltage intervals where the exponent $n$ in Eq.~\ref{calcn} are
nearly constant, corresponding values of $n$ are indicated in the
figure.} \label{alpha-V_79}
\end{figure}

The explanation of noise decrease with increasing bias can be straightforwardly based on a scenario in which bias causes changes in the symmetry of two level fluctuators in the ensemble responsible for generating the observed noise. Elementary fluctuators in intrinsic tunnel junctions can be associated, for example, with charge traps within tunnel barriers and two distinct states of a fluctuator with empty and loaded state of the trap, or alternatively, with charged oxygen vacancies jumping between two sites. Electrostatic field of the charge trap modulates the height of the tunnel energy barrier and, consequently, the resistivity of the junction, according to the trap occupancy state.

An alternative mechanism consists in bias induced changes to the regions where relaxation of tunneling charges takes place. At low bias, when elastic tunneling dominates, the transmitted electrons relax energy in conducting regions through interactions with bosonic degrees of freedom, whereas for inelastic tunneling the relaxation occurs inside the insulating barriers. If in the latter case the activation energy of defects within the barrier is higher than that of the conducting regions, then the  noise will decrease with increasing intensity of the inelastic tunneling events. The observed bias induced changes in $P(\Delta )$ distribution with changing bias do allow, at this stage, to choose one from the above discussed scenarios. One has however to note that the energy threshold corresponding to the onset on nonequlibrium noise at threshold bias of 200 mV is much lower than the activation  energy of oxygen vacancies in perovskites. At the same time we observe that this is the energy level of the Jahn-Teller distortions in manganites. \cite{Jooss}

\section{Conclusions}
In conclusion, we have investigated conductivity fluctuations in
current biased La$_{0.82}$Ca$_{0.18}$MnO$_{3}$ single crystals in
a wide range of bias current and temperature, a very good model
object for studying the noise processes in inhomogeneous
conducting media since different magnetic and electric transport
properties within a temperature interval from the liquid-nitrogen
temperature up to ambient one. The observed noise retains $1/f$
spectra in entire experimentally explored range of bias and
temperature despite pronounced changes in the dissipation
mechanism and magnetic state that occur in the same range. The
voltage noise measured at constant current bias $I$ was found to
have characteristics of a quasi-equilibrium modulation noise with
intensity scaling as $I^2$, indicating that the conductivity
fluctuations are only revealed by the current flow and not
stimulated or modified by it. Only at liquid nitrogen temperatures
the observed $1/f$ noise behaves like nonequilibrium one and its
intensity not only does not scale with the square of the bias but
even decreases locally with increasing bias.

We have verified that in entire experimentally investigated range
the noise fulfils the Dutta-Horn model frequency-temperature
reciprocity condition. We have concluded therefore that noise in
our system arises from a superposition of thermally activated
processes with broad distribution of activation energies. However,
basing this conclusion just on the consistency between temperature
and frequency dependence of the noise has to be taken with the
grain of salt since the noise model based on serious kinetics of a
random walk in random potential (RWRP), a model that in principle
could apply well to our 0.18 LCMO sample, produces
frequency-temperature reciprocity relation identical to Eq.
(\ref{DH}). Only in-depth analysis of non-Gaussian components
of the noise, provided they exist, enables one to distinguish
between the reality of two models.\cite{weiss-rew} We did not
observe any non-Gaussian  noise components in the experiment and
are unable to distinguish between the two models.

Nevertheless, we have tentatively accepted the conclusion that the
noise indeed arises from the DH mechanism and inferred the
distributions of activation energies in the elementary
fluctuations ensembles by analyzing the power spectral density of
the noise within the framework of the DH approach. We have found
that the appearance of a nonequilibrium noise is directly linked
with the onset of bias dependence of the distribution of
activation energies $P(E)$. We have found clear correlations
between the changes of the energy distribution and the power
exponent of the $I-V$ curves. Changes in the power exponent, i.e.,
the hallmarks of the changes in the dissipation mechanism, are
correlated with changes in the noise behavior. We have discussed
two possible scenarios: (i) a direct influence of the electric
field on fluctuators which in this case should be charged and (ii)
an effect of the spatial change of the region where the electron
energy dissipates during its traveling across an inhomogeneous
sample. Both models are appropriate in our case and we are
performing additional studies in order to reveal the most probable
one.

The results obtained in the work open a perspective for applying
the noise tool to investigate the nature and physics involved in
metastable resistivity states which can be enforced in low-doped
manganites by applying strong pulsed electric field/current at low
temperature.~\cite{PRB018,PRB020} Our first experiments have
confirmed this assumption and their analysis will be published
elsewhere.

\acknowledgments This research was supported by the Israeli Science Foundation administered by the Israel Academy of Sciences and Humanities (grant 754/09).


\begin{thebibliography}{99}

\bibitem{DH} P. Dutta and P. M. Horn,  Rev. Mod. Phys. {\bf 53}, 497 (1981).
\bibitem{weiss-rew} M. B. Weissman, Rev. Mod. Phys. {\bf 60}, 537 (1998).
\bibitem{weiss-tool} M. B. Weissman, Annu. Rev. Mater. Sci. {\bf 26}, 395 (1996).
\bibitem{raquet} B. Raquet, J. M. P. Coy, and S. von Moln´ar, Phys. Rev. B {\bf 59}, 12435 (1999).
\bibitem{raquetPRL} B. Raquet, A. Anane, S. Wirth, P. Xiong, and S. von Moln´ar, Phys. Rev. Lett. {\bf 84}, 4485 (2000).
\bibitem{podz}V. Podzorov, M. Uehara, M. E. Gershenson, T. Y. Koo, and S-W. Cheong,  Phys. Rev. B {\bf 61}, R3784 (2000).
\bibitem{alers} G. B. Alers, A. P. Ramirez, and S. Jin, Appl. Phys. Lett. {\bf 68}, 3644 (1996).
\bibitem{ahn} K. H. Ahn, T. Lookman, A. R. Bishop,  Nature {\bf 428}, 401 (2004).
\bibitem{ourAPL} X. D. Wu, B. Dolgin, G. Jung, V. Markovich, Y. Yuzhelevski, M. Belogolovskii, Ya. M. Mukovskii, Appl. Phys. Lett. {\bf 90}, 242110 (2007).
\bibitem{reut} P. Reutler, A. Bensaid, F. Herbstritt, C. Hofener, A. Marx, and R. Gross, Phys. Rev. B {\bf 62}, 11619 (2000).
\bibitem{palani} A. Palanisami, R. D. Merithew, M. B. Weissman, M. P. Warusawithana, F. M. Hess, and J. N. Eckstein,  Phys. Rev. B {\bf 66}, 092407 (2002).
\bibitem{rana} D. S. Rana, M. Ziese, S. K. Malik,  Phys. Rev. B {\bf 74}, 094406 (2006).
\bibitem{phillip} J. B. Philipp,  L. Alff, A. Marx, and R. Gross, Phys. Rev. B {\bf 66}, 224417 (2002).
\bibitem{barone} C. Barone, C. Adamo, A. Galdi, P. Orgiani, A. Yu. Petrov, O. Quaranta, L. Maritato, and S.
Pagano,  Phys. Rev. B {\bf 75}, 174431 (2007).
\bibitem{nowak} A. Gokce, E. R. Nowak, S. H. Yang and S. S. P. Parkin, J. Appl. Phys. {\bf 99}, 08A906 (2006).
\bibitem{crystal} D. A Shulyatev, A. A. Arsenov, S. G. Karabashev, Ya. M. Mukovskii, J. Crys. Growth, {\bf 198/199}, 511 (1999).
\bibitem{PRB018} Y. Yuzhelevski, V. Markovich, V. Dikovsky, G. Gorodetsky, G. Jung, D. A. Shulyatev, and Ya. M. Mukovskii, Phys. Rev. B {\bf  64}, 224428 (2001).
\bibitem{zygulski} G. P. Zhigalski, Phys. Usp., {\bf 46}, 449 (2003).
\bibitem{imry} E. Pytte and Y. Imry, Phys. Rev. B {\bf 35}, 1465 (1987).
\bibitem{Wolf}  E. L. Wolf, Principles of Electron Tunneling Spectroscopy (Oxford University Press, Oxford, 1985).
\bibitem{Maximov}  E. G. Maksimov and O. V. Dolgov, Phys. Usp. {\bf 50}, 933 (2007).
\bibitem{Adams}  C. P. Adams, J. W. Lynn, V. N. Smolyaninova, A. Biswas, R. L. Greene, W. Ratcliff, S.-W. Cheong, Y. M. Mukovskii, D. A. Shulyatev, Phys. Rev. B {\bf 70}, 134414 (2004).
\bibitem{kirtley} J. R. Kirtley, D. J. Scalapino, Phys. Rev. Lett. {\bf 65}, 799 (1990).
\bibitem{bel} M. A. Belogolovskii, Yu. F. Revenko, A. Yu. Gerasimenko, V. M. Svistunov, E. Hatta, G. Plitnik, V. E. Shaternik, and E. M. Rudenko, Low Temp. Phys. {\bf 28}, 391 (2002).
\bibitem{GM} L. I. Glazman and K. A. Matveev, Sov. Phys. JETP {\bf 67}, 1276 (1988).
\bibitem{Beasley}  Y. Xu, D. Ephron, and M. Beasley, Phys. Rev. B {\bf 52}, 2843 (1995).
\bibitem{gross} J. Klein, C. Hofener, S. Uhlenbruck, L. Al, B. Buchner, and R. Gross, Europhys. Lett. {\bf 47}, 371 (1999).
\bibitem{bertina} K. B. Chashka, B. Fisher, J. Genossar, L. Patlagan, G. M. Reisner, and E. Shimshoni, Phys. Rev. B {\bf 63}, 064403 (2001).
\bibitem{CEJP}  M. Belogolovskii, Cent. Eur. J. Phys. {\bf 7}, 304 (2009).
\bibitem{Machlup} S. Machlup, J. Appl. Phys. {\bf 25}, 341 (1954).
\bibitem{PRB020}V. Markovich, G. Jung, Y. Yuzhelevski, G. Gorodetsky, A. Szewczyk, M. Gutowska, D. A. Shulyatev and Ya. M. Mukovskii, Phys. Rev. B {\bf 70}, 064414 (2004).
\bibitem{Jooss} Ch. Jooss, J. Hoffmann, J. Fladerer, M. Ehrhardt, T. Beetz, L. Wu, and Y. Zhu, Phys. Rev. B {\bf 77}, 132409 (2008).
\end{thebibliography}
\end{document}